\documentstyle[]{article}
\title{ Only hybrid anyons can exist  \\
        in broken symmetry phase of  \\
        nonrelativistic $[U(1)]^{2}$ Chern-Simons theory }
\author{Jacek Dziarmaga  \\
        Jagellonian University, Institute of Physics,  \\
        Reymonta 4,30-059 Krak\'ow, Poland
        \thanks{Permanent address. E-mail: ufjacekd@ztc386a.if.uj.edu.pl}\\
                        and                            \\
        Institute for Theoretical Physics, University of Utrecht \\}
\date{29 April 1994}
\begin{document}
\maketitle

    \begin{abstract}
    We present two examples of parity-invariant $[U(1)]^{2}$
Chern-Simons-Higgs models with spontaneously broken symmetry. The models
possess topological vortex excitations. It is argued that the smallest
possible flux quanta  are composites of one quantum of each type $(1,1)$.
These hybrid anyons will dominate the statistical properties near the
ground state. We analyse their statistical interactions and find out that
unlike in the case of Jackiw-Pi solitons there is short range magnetic
interaction which can lead to formation of bound states of hybrid anyons.
In addition to mutual interactions they possess internal structure
which can lead upon quantisation to discrete spectrum of energy levels.
    \end{abstract}
\vspace{2\baselineskip}
hep-th 9404182\\[\baselineskip]
  Chern-Simons interaction does not necessarily lead to parity breaking
as it was recently pointed out \cite{wilczek,hagen}. One can construct
parity-invariant theories with an even number of C-S fields.
Field theoretical models of this type were analysed
\cite{5,cs2} as well as microscopical theories \cite{ezawa2}. The topic is
interesting
because high-$T_{c}$ superconductors seem to be layered systems
but there is no experimental indication of parity breaking.
In this paper
we would like to present two Bogomol'nyi type models in which the ground
state will be dominated by topological vortices with winding numbers
$(1,1)$. Vortices of this type are anyons in spite of the fact that the
theory is parity invariant. This ground state is a solution with
spontaneously broken P-invariance. We also analyse their statistical
interactions and come to the conclusion that they can form bound states
due to short range magnetic trapping.

\section{The model with condensate induced by external magnetic field
                                and normalisability of splitting modes}

Let us consider Lagrangian density of the double layer system
\begin{eqnarray}\label{10}
L&=&\kappa\varepsilon^{\mu\nu\lambda}
                     a^{(1)}_{\mu}\partial_{\nu}a^{(2)}_{\lambda} +
\frac{1}{2}i(\psi^{\star}D_{t}\psi-\psi D_{t}\psi^{\star})
                        -\frac{1}{2}D_{k}\psi^{\star}D_{k}\psi  \nonumber \\
 &+&\frac{1}{2}i(\phi^{\star}D_{t}\phi-\phi D_{t}\phi^{\star})
            -\frac{1}{2}D_{k}\phi^{\star}D_{k}\phi-U(\psi,\phi)
\end{eqnarray}
where $\kappa$ is choosen to be positive for definiteness. The covariant
derivatives and the potential are
\begin{eqnarray}\label{20}
D_{\mu}\psi=\partial_{\mu}\psi-ia^{(1)}_{\mu}\psi-iA_{\mu}\psi\;\;,\;\;
D_{\mu}\phi=\partial_{\mu}\phi-ia^{(2)}_{\mu}\phi-iA_{\mu}\phi\;\;,\nonumber \\
U=\frac{1}{\kappa}\mid\psi\mid^{2}\mid\phi\mid^{2}
          -\frac{\mid B_{ext} \mid}{2}(\mid\psi\mid^{2}+\mid\phi\mid^{2})\;\;.
\end{eqnarray}
This is a simplified version of the Lagrangian in \cite{ezawa2} and of the
nonrelativistic model in \cite{5}. Chemical potentials are proportional
to $\mid B_{ext}\mid$ and the mixed term of the fourth order can be
regarded as a renormalisation counterterm \cite{ezawa1,ezawa2}. Coulomb
interaction \cite{ezawa1,ezawa2}
is neglected at the beginning. The chemical potentials can depend
on $B^{ext}$ also in some other way, say $\mu(B^{ext})$. This model is
Bogomol'nyi type of theory only at such a special value (values)
of external magnetic field at which
$\mu(B^{ext})=\frac{1}{2}\mid B^{ext} \mid$.
We have restricted to the case with purely mutual statistical
interaction to provide the paper with greater clarity. The external potentials
$A_{\mu}$ were introduced to provide the uniform external magnetic field
background
\begin{equation}\label{30}
  B^{ext}=-\varepsilon_{mn}\partial_{m}A_{n}^{ext} \;\;,\;\;
  A_{0}^{ext}=0 \;\;.
\end{equation}
Variation of the Lagrangian with respect to $a^{(I)}_{0}$ , with
$I=1,2$, leads to Gauss' laws
\begin{equation}\label{40}
  \kappa B_{1}=\phi^{\star}\phi \;\;,\;\;
  \kappa B_{2}=\psi^{\star}\psi \;\;,
\end{equation}
which should be regarded as constraints with Lagrange multipliers:
$a_{0}^{(I)}$. With such a constraint the Hamiltonian density reads
\begin{equation}\label{50}
  H=\frac{1}{2}D_{k}\psi^{\star}D_{k}\psi+
    \frac{1}{2}D_{k}\phi^{\star}D_{k}\phi+U \;\;.
\end{equation}
Now we apply the Bogomol'nyi trick
\begin{equation}\label{60}
  D_{k}\psi^{\star}D_{k}\psi=\mid(D_{1}+iD_{2})\psi\mid^{2}
       \stackrel{-}{+}B_{1}\rho_{1}\stackrel{-}{+}B^{ext}\rho_{1}
       \stackrel{-}{+}\nabla\times \vec{J}_{1} \;\;.
\end{equation}
With this decomposition and with the use of the Gauss' laws one obtains
after neglect of boundary terms, for $B^{ext}<0$
\begin{equation}\label{70}
 H=\frac{1}{2}\mid(D_{1}+iD_{2})\psi\mid^{2}+
   \frac{1}{2}\mid(D_{1}+iD_{2})\phi\mid^{2} \;\;,
\end{equation}
which is positively definite. This Hamiltonian is minimised by fields
satisfying following equations
\begin{equation}\label{80}
  D_{+}\psi \;\;\;,\;\;\; D_{+}\phi=0 \;\;,
\end{equation}
together with Gauss' laws. Solutions to the above equations are also
solutions to the Euler-Lagrange equations of the model provided that Lagrange
multipliers are taken equal to
\begin{equation}\label{100}
  a^{(1)}_{0}=\frac{1}{2\kappa}\rho_{2} \;\;\;,\;\;\;
  a^{(2)}_{0}=\frac{1}{2\kappa}\rho_{1} \;\;,
\end{equation}
{}From the self-duality equations (\ref{80}) we obtain
\begin{equation}\label{110}
   a^{(I)}_{k}=\frac{1}{2}\varepsilon_{kl}\partial_{l}\ln\rho_{I}+
               \partial_{k}\omega_{I}-A_{k}^{ext} \;\;,\;\;
   (\;1\;\leftrightarrow\;2\;) \;\;,\;\;
   I=1,2 \;\;,
\end{equation}
where $\omega_{1,2}$ are phases of the fields $\psi$ and $\phi$ respectively.
Substitution to the Gauss' laws (\ref{40}) yields
\begin{equation}\label{120}
  \nabla^{2}\ln\rho_{1}=\frac{2}{\kappa}(\rho_{2}-\rho_{0})+
  2\varepsilon_{mn}\partial_{m}\partial_{n}\omega_{1}\;\;\;,\;\;\;
  (\;1\;\leftrightarrow\;2\;) \;\;\;,\;\;\; \rho_{0}=\kappa\mid
B_{ext}\mid\;\;.
\end{equation}
Assuming the phases of the Higgs fields to be of the form
\begin{equation}\label{130}
  \omega_{I}=\sum_{p_{I}}\Theta(\vec{x}-\vec{R}_{p_{I}}) \;\;,\;\;
  p_{I}=1,...,n_{I} \;\;,
\end{equation}
which fixes the gauge $\partial_{k}a_{k}^{(I)}=0$, we get
\begin{eqnarray}\label{140}
  \nabla^{2}\ln\rho_{1}&=&\frac{2}{\kappa}(\rho_{2}-\rho_{0})+
  4\pi\sum_{p_{1}} \delta^{(2)}(\vec{x}-\vec{R}_{p_{1}}) \;\;, \nonumber \\
  \nabla^{2}\ln\rho_{2}&=&\frac{2}{\kappa}(\rho_{1}-\rho_{0})+
  4\pi\sum_{p_{2}} \delta^{(2)}(\vec{x}-\vec{R}_{p_{2}}) \;\;.
\end{eqnarray}
In the special case of $\rho_{1}=\rho_{2}\equiv\rho$ and outside of the
singular points we will obtain
\begin{equation}\label{160}
  \nabla^{2}\ln\rho=\frac{2}{\kappa}(\rho-\rho_{0}) \;\;.
\end{equation}
This equation is known to possess static multivortex solutions \cite{taubes}.
The separate vortices are in fact composites of identical vortices
of two different types sitting on top of each another. Natural question
arises whether we can expect also existence of separate vortices of diferent
types. Now we will explicitely count the number of normalisable zero modes
around the static (1,1) solution. It will appear that there are only two
translational modes. The modes which could correspond to splitting of
(1,1) vortex into separate (1,0) and (0,1) vortices are not normalisable
and thus would lead to unbounded rise of charge with respect to background.
We will consider special configuration of coinciding vortices with
vorticity $(n,n)$.

   Let the unperturbed solution be of the form
\begin{equation}\label{170}
  \psi=\phi=f(r)e^{in\theta} \;\;.
\end{equation}
Now we take perturbations of the phases of the Higgs field to be
$\alpha_{1},\alpha_{2}$ and those of the moduli: $fh_{1}$ , $fh_{2}$.
\begin{eqnarray}\label{180}
  \psi+\delta\psi=f(r)[1+h_{1}(r,\theta)]
                                  e^{in\theta+i\alpha_{1}(r,\theta)} \;\;,
                                                                 \nonumber \\
  \phi+\delta\phi=f(r)[1+h_{2}(r,\theta)]
                                  e^{in\theta+i\alpha_{2}(r,\theta)} \;\;.
\end{eqnarray}
Linearisation of the self-dual equations (\ref{80}) with respect
to perturbations of the Higgs fields and those of gauge potentials
$c_{k}^{(I)}$ , $I=1,2$, yields
\begin{eqnarray}\label{190}
   c_{\theta}^{(I)}=\partial_{r}h_{I}-\frac{1}{r}\partial_{\theta}\alpha_{I}
                                                            \;\;, \nonumber \\
   c_{r}^{(I)}=-\frac{1}{r}\partial_{\theta}h_{I}-\partial_{r}\alpha_{I}
\end{eqnarray}
for $I=1,2$. Once the perturbations of the Higgs field are known,
$c_{k}^{(I)}$ can be calculated from the above equations. To have a unique
solution we have to fix the gauge
\begin{equation}\label{200}
  \partial_{k}c_{k}^{(I)}\equiv
    \partial_{r}c_{r}^{(I)}+\frac{1}{r}c_{r}^{(I)}+
    \frac{1}{r}\partial_{\theta}c_{\theta}^{(I)}=0 \;\;,\;\; I=1,2\;.
\end{equation}
Upon substitution of (\ref{190}) to this gauge condition we will obtain
\begin{equation}\label{210}
  \nabla^{2}\alpha_{1}=0 \;\;\;,\;\;\; \nabla^{2}\alpha_{2}=0 \;\;.
\end{equation}
Similar substitution of (\ref{190}) to Gauss' laws (\ref{40}) will lead to
\begin{equation}\label{220}
  \nabla^{2}h_{1}=-\frac{2}{\kappa}f^{2}(r)h_{2} \;\;,\;\;
  \nabla^{2}h_{2}=-\frac{2}{\kappa}f^{2}(r)h_{1} \;\;.
\end{equation}
Now let us consider the following most general Ansatz
\begin{eqnarray}\label{230}
  h_{1}(r,\theta)&=&\sum_{k=1}^{n}
   [\xi_{k}^{1}P_{k}(r)\cos k\theta+\xi_{k}^{2}Q_{k}(r)\sin k\theta] \;\;,
                                                             \nonumber \\
  h_{2}(r,\theta)&=&\sum_{k=1}^{n}
   [\lambda_{k}^{1}P_{k}(r)\cos k\theta+\lambda_{k}^{2}Q_{k}(r)\sin k\theta]
                                                                     \;\;,
\end{eqnarray}
with $4n$ real parameters. This Ansatz leads to following two sets
of equations
\begin{eqnarray}\label{240}
  \xi_{k}^{1}[\triangle_{k}P_{k}(r)]&=&
                   [-\frac{2}{\kappa}f^{2}(r)P_{k}(r)]\lambda_{k}^{1} \;\;,
                                                             \nonumber \\
  \lambda_{k}^{1}[\triangle_{k}P_{k}(r)]&=&
                   [-\frac{2}{\kappa}f^{2}(r)P_{k}(r)]\xi_{k}^{1} \;\;,
\end{eqnarray}
and
\begin{eqnarray}\label{250}
  \xi_{k}^{2}[\triangle_{k}Q_{k}(r)]&=&
                   [-\frac{2}{\kappa}f^{2}(r)Q_{k}(r)]\lambda_{k}^{2} \;\;,
                                                             \nonumber \\
  \lambda_{k}^{2}[\triangle_{k}Q_{k}(r)]&=&
                   [-\frac{2}{\kappa}f^{2}(r)Q_{k}(r)]\xi_{k}^{2} \;\;,
\end{eqnarray}
where the laplacians are
\begin{equation}\label{260}
  \triangle_{k}=
   \frac{d^{2}}{dr^{2}}+\frac{1}{r}\frac{d}{dr}-\frac{k^{2}}{r^{2}} \;\;.
\end{equation}
These equations do not lead to contradiction for nonzero
$\lambda$ , $\xi$ only if
\begin{equation}\label{270}
  \frac{\xi_{k}^{1}}{\lambda_{k}^{1}}\equiv\sigma_{k}^{1}=\stackrel{+}{-}1
                         \;\;\;and\;\;\;
  \frac{\xi_{k}^{2}}{\lambda_{k}^{2}}\equiv\sigma_{k}^{2}=\stackrel{+}{-}1
                                                                       \;\;,
\end{equation}
where the choices of signs are independent. Now we have only two equations
\begin{eqnarray}\label{280}
  P_{k}^{\prime\prime}+\frac{1}{r}P_{k}^{\prime}+
    [\frac{2\sigma^{1}_{k}}{\kappa}f^{2}-\frac{k^{2}}{r^{2}}]P_{k}=0 \;\;,
                                                              \nonumber \\
  Q_{k}^{\prime\prime}+\frac{1}{r}Q_{k}^{\prime}+
    [\frac{2\sigma^{2}_{k}}{\kappa}f^{2}-\frac{k^{2}}{r^{2}}]Q_{k}=0 \;\;.
\end{eqnarray}
Asymptotically at $r\rightarrow\infty$ they become Bessel or modified Bessel
equations dependent on whether given $\sigma^{I}_{k}$ is negative
or positive, since $f^{2}(\infty)=\rho_{0}$. Thus the only normalisable
zero modes are those
with $\sigma_{k}^{1}=\sigma_{k}^{2}=1$ or in terms of the parameters
\begin{equation}\label{290}
  \xi_{k}^{1}=\lambda_{k}^{1} \;\;\; and \;\;\;
  \xi_{k}^{2}=\lambda_{k}^{2} \;\;.
\end{equation}
With this condition we have effectively only one equation
\begin{equation}\label{300}
  P_{k}^{\prime\prime}+\frac{1}{r}P_{k}^{\prime}+
         [\frac{2}{\kappa}f^{2}-\frac{k^{2}}{r^{2}}]P_{k}=0 \;\;\;,\;\;\;
  Q_{k}=P_{k} \;\;.
\end{equation}
The asymptotically vanishing solution possesses singularity
at the origin, which can be normalised as
\begin{equation}\label{310}
  P_{k}(r)\sim -\frac{1}{r^{k}} \;\;\;,\;\;\;r\rightarrow 0\;\;.
\end{equation}
With this asymptotics we can choose such a solution to eq.(\ref{210}) that
gauge potentials expressed as in eq.(\ref{190}) in terms of perturbations
of the moduli and phase of the Higgs field are regular at the origin.
This choice appears to be unique
\begin{equation}\label{320}
  \alpha=\alpha_{1}=\alpha_{2}=\frac{1}{r^{k}}
              [\lambda^{1}_{k}\sin k\theta-\lambda^{2}_{k}\cos k\theta] \;\;.
\end{equation}
After we expand the perturbed Higgs fields around the origin we will
obtain up to terms linear in $\lambda$
\begin{equation}\label{330}
  \psi+\delta\psi=\phi+\delta\phi\approx
      (z^{n}-\sum_{k}\lambda_{k}z^{k}) \;\;\;,\;\;\;
      \lambda_{k}=\lambda_{k}^{1}+i\lambda_{k}^{2} \;\;.
\end{equation}
Now we can clearly see that the only normalisable zero modes are those which
correspond to splitting of the coincident $(n,n)$ configuration into a set
of hybrid vortices of the type $(p,p)$. Splitting into vortices of the type
$(p,q)$ with $p\neq q$ is forbidden by charge conservation.

   We have analysed splitting modes only within the framework of self-dual
equations. This analysis shows that the coincident (1,1) configuration
sits at the bottom of potential well. The question now is whether the
energy of a pair of (1,0) and (0,1) vortices is higher or the same as that
of coincident solution. If the latter is the case it would mean that
a pair of separate vortices is separated from the coincident configuration
by a potential barrier and otherwise degenerate. At low temperatures we
would have two coexistent phases of vortex condensates. We can give a simple
argument to rule out the latter possibility. Let us analyse similarly as
in \cite{5} asymptotic properties of equations (\ref{140}). We define small
fluctuations around asymptotic values by
\begin{equation}\label{n1}
  g_{1}=\frac{2}{\kappa}(\rho_{1}-\rho_{0}) \;\;\;,\;\;\;
  g_{2}=\frac{2}{\kappa}(\rho_{2}-\rho_{0}) \;\;.
\end{equation}
Equations (\ref{140}) when linearised with respect to $g_{1}$ and $g_{2}$
take the form
\begin{eqnarray}\label{n2}
  \nabla^{2}g_{1}&=&\alpha g_{2} \;, \nonumber \\
  \nabla^{2}g_{2}&=&\alpha g_{1} \;\;\;,\;\;\;
  \alpha=\frac{2\rho_{0}}{\kappa}>0 \;.
\end{eqnarray}
We can remove $\alpha$ by rescaling coordinates and then diagonalise
this set of equations
\begin{eqnarray}\label{n3}
  \nabla^{2}(g_{1}+g_{2})= + (g_{1}+g_{2}) \;, \nonumber \\
  \nabla^{2}(g_{1}-g_{2})= - (g_{1}-g_{2}) \;.
\end{eqnarray}
Nontrivial asymptotic solutions are given by asymptotics of Bessel functions.
$(g_{1}+g_{2})$ vanishes exponentialy. An asymptotic decay of $(g_{1}-g_{2})$
is much slower so only the trivial solution $g_{1}=g_{2}$ is compatible
with finiteness of $U(1)$ charges. This argument excludes existence of
zero energy vortex with winding numbers (1,0) or (0,1). Thus the energy
of the widely separated pair of (1,0) and (0,1) vortices must be higher
then the energy of the hybrid anyon with winding numbers (1,1).

    Thus we can expect that at low temperatures the system will be
dominated by minimal composites with the winding numbers $(1,1)$.

\section{Model of statistical interaction with uniform background}

  Another model with dominance of the ground state by hybrid anyons is
a direct extension of the U(1) model proposed by Barashenkov and Harin
\cite{igor}
\begin{eqnarray}\label{1010}
L&=&\kappa\varepsilon^{\mu\nu\lambda}
                            a^{(1)}_{\mu}\partial_{\nu}a_{\lambda}^{(2)}
             +\frac{1}{2}i(\psi^{\star}D_{t}\psi-\psi D_{t}\psi^{\star})
             -\frac{1}{2}D_{k}\psi^{\star}D_{k}\psi            \nonumber \\
 &+&\frac{1}{2}i(\phi^{\star}D_{t}\phi-\phi D_{t}\phi^{\star})
  -\frac{1}{2}D_{k}\phi^{\star}D_{k}\phi-V(\phi,\psi)
  -q a_{0}^{(1)}-q a_{0}^{(2)} \;\;,
\end{eqnarray}
with potential of the form
\begin{equation}\label{1020}
  V(\phi,\psi)=\frac{1}{\kappa}(\rho_{1}-q)(\rho_{2}-q) \;\;.
\end{equation}
Gauss' laws in this case read
\begin{equation}\label{1030}
  \kappa B_{1}=(\rho_{2}-q)  \;\;\;,\;\;\;
  \kappa B_{2}=(\rho_{1}-q)  \;\;.
\end{equation}
The Hamiltonian after Bogomol'nyi decomposition is
\begin{equation}\label{1040}
  H=\frac{1}{2}\mid D_{+}\psi \mid^{2} +
    \frac{1}{2}\mid D_{+}\phi \mid^{2} +
    \frac{1}{2}q(B_{1}+B_{2})   \;\;.
\end{equation}
The energy is bounded from below by
\begin{equation}\label{1050}
  E\geq 2\pi(n_{1}+n_{2}) \;\;.
\end{equation}
This lower bound is saturated by fields satisfying
\begin{equation}\label{1060}
  D_{+}\psi=0 \;\;\;,\;\;\; D_{+}\phi=0 \;\;.
\end{equation}
The Lagrange multipliers have to be
\begin{equation}\label{1070}
  a_{0}^{(1)}=\frac{1}{2\kappa}(\rho_{2}-q)  \;\;\;,\;\;\;
  a_{0}^{(2)}=\frac{1}{2\kappa}(\rho_{1}-q)  \;\;.
\end{equation}
Further analysis similar as in section 1 shows that also in this case
vortex with winding numbers $(n,n)$ can decay only into vortices
of the type $(p,p)$ and not in those of the type $(p,q)$.
Once again we are lead to conclusion that hybrid anyons will dominate
statistically at low temperatures.

\section{Statistics and magnetic interactions of hybrid anyons}

   Now we will assume that the ground state of the theory is dominated by
elementary hybrid anyons of the type $(1,1)$. We will apply Manton's
\cite{manton}
approximation to investigate their statistical properties. Namely we assume
that a good approximation to configuration of slowly moving vortices
are static multivortex solutions with their parameters promoted to the role
of time dependent collective coordinates. This method was first applied
in the context of relativistic Chern-Simons vortices by Kim and Min
\cite{kimmin}.
In our Lagrangians (\ref{10}) or (\ref{1010})
we have first order time derivatives so this prescription can be
directly applied only to obtain those terms of the effective Lagrangian
which are
linear in time derivatives of the parameters. The part of the effective
Lagrangian linear in velocities reads
\begin{equation}\label{510}
  L^{(1)}_{eff}=\int d^{2}x [
          -\kappa\varepsilon^{mn}a_{m}^{(1)}\partial_{t}a_{n}^{(2)}
          -\rho_{1}\partial_{t}\omega_{1}-\rho_{2}\partial_{t}\omega_{2}]\;\;,
\end{equation}
The first term gives no contribution since the gauge potentials satisfy the
Coulomb gauge $\partial_{k}a_{k}^{(I)}=0$ at every moment of time. Now we
restrict to the dominant multivortex solution, which satisfies $\psi=\phi$,
$a_{\mu}^{(1)}=a_{\mu}^{(2)}$,
\begin{equation}\label{520}
  L^{(1)}_{eff}=-2\int d^{2}x\;(\rho\partial_{t}\omega)=
                       -2\kappa\int d^{2}x\; B\partial_{t}\omega \;\;,
\end{equation}
where we have made use of the Gauss' laws (\ref{40}) or (\ref{1040}).
With the form of the phase
\begin{equation}\label{530}
  \dot{\omega}=\frac{d}{dt}\sum_{p}\Theta(\vec{x}-\vec{R}_{p})=
                   \sum_{p}\varepsilon_{ij}\dot{R}^{i}_{p}\partial_{j}
                                        \ln\mid\vec{x}-\vec{R}_{p}\mid
\end{equation}
and some integration by parts we get the form of effective Lagrangian useful
for further discussion.
\begin{equation}\label{540}
  L^{(1)}_{eff}=-4\pi\kappa\sum_{p}\dot{R}^{i}_{p}a_{i}(\vec{R}_{p}) \;\;,
\end{equation}
which is expressed through the gauge field at the cores of vortices and their
velocities. Now we can easily get the orbital part of the angular momentum
at vanishing velocities
\begin{equation}\label{550}
 J_{orb}=\sum\varepsilon_{ij}R^{i}_{p}
         \frac{\partial L^{(1)}_{eff}}{\partial \dot{R}^{j}_{p}}=
         -4\pi\kappa\sum_{p}R^{i}_{p}\varepsilon_{ij}a_{j}(\vec{R}_{p}) \;\;.
\end{equation}
It is worth noticing that the effective Lagrangian (\ref{540}) takes the form
of the coupling of the point particle current to external gauge field
defined on the moduli space. Due to this coupling vortex at $\vec{R}_{p}$
feels the magnetic field
\begin{equation}\label{560}
B_{eff}(\vec{R}_{p})=
   -4\pi\kappa\varepsilon_{ij}\partial_{i}a_{j}(\vec{R}_{p})  \;\;\;,\;\;\;
                \partial_{i}\equiv\frac{\partial}{\partial R^{i}_{p}} \;\;.
\end{equation}
Now let us consider configuration of two vortices each of the type $(1,1)$
in the center of mass frame at actual positions $+\vec{R}$ and
$-\vec{R}$. The orbital momentum reads
\begin{equation}\label{570}
J_{orb}=-4\pi\kappa[R^{i}\varepsilon_{ij}v_{j}(\vec{R})-
                    R^{i}\varepsilon_{ij}v_{j}(-\vec{R})]=
        -8\pi\kappa Rv_{\theta}(R) \;\;,
\end{equation}
where the last equality holds due to the symmetry of the configuration.
It is easy to see that
\begin{equation}\label{580}
 \frac{1}{R}\frac{dJ(R)}{dR}=
  -8\pi\kappa[\frac{v_{\theta}}{R}+\frac{dv_{\theta}(R)}{dR}]=
   B_{eff}(+\vec{R})+B_{eff}(-\vec{R})=B_{eff}(R) \;\;.
\end{equation}
Thus if the spin dependence of two-vortex static configuration on their
distance were known the magnetic interaction could be extracted due to the
above formula. The formula makes use of the global properties of the solution
so it is not very sensitive to numerical errors. The total angular momentum
$J$ may differ from its orbital part only by an additive normalisation
constant.

  Explicite calculation of the linear part of effective Lagrangian
from eq.(\ref{520}) leads for well separated vortices to
\begin{equation}\label{590}
  L^{(1)}_{eff}=4\pi\kappa\sum_{p>q}\dot{\Theta}(\vec{R}_{p}-\vec{R}_{q})
\;\;,
\end{equation}
Short range interactions are more complicated. We can promote $\lambda$-s
in eq.(\ref{230}) in the case $n=2$ , $k=2$ to the role of collective
coordinates. The effective Lagrangian up to leading terms in $\lambda$ can be
evaluated as
\begin{equation}\label{600}
  L^{(1)}_{eff}=b_{eff}\Lambda^{2}\dot{\omega} \;\;\;,\;\;\;
  b_{eff}=\pi \int_{0}^{\infty} r\;dr\;\frac{f^{2}P_{2}}{r^{2}}  \;\;,
\end{equation}
where $\lambda_{2}^{1}+i\lambda_{2}^{2}=\Lambda e^{i\omega}$. Positions
of vortices are complex square roots of $\lambda$. If they are
$+\vec{R}$ and $-\vec{R}$ then in polar coordinates
$R^{2}e^{2i\Theta}=\Lambda e^{i\omega}$ and
\begin{equation}\label{610}
  L^{(1)}_{eff}=2b_{eff}R^{4}\dot{\Theta} \;\;,
\end{equation}
which describes the short range asymptotics of magnetic interaction.

   So far we have considered only statistical and magnetic interactions
of hybrid anyons as a whole. The hybrid anyon can be thought of as a pair
of (1,0) and (0,1) vortices trapped in a potential well. In \cite{cs2}
it was shown that there are mutual magnetic short range interactions
between the different types of vortices. In the center of mass frame
the internal dynamics of single hybrid anyon reduces to that of
short range oscillator in external magnetic field. Quantum mechanically
we can expect an interesting discrete spectrum.

\section{Conclusions}

  In both cases analised in this paper we are lead to the conclusion that
the dominant flux quanta are vortices with winding numbers $(1,1)$.
In the second model they also have smallest energy thanks to
the Bogomol'nyi lower bound which they saturate and thus will dominate
statistically. In the case of the first model all solutions are degenerate
in energy. Degeneracy can be removed by introduction of Coulomb interaction
which will rise the energy of excitations with bigger winding number
equivalent to bigger charges.

  The short range remnant of statistical interaction is rather a magnetic
intertaction unlike in the case of Jackiw-Pi solitons investigated by
the same method \cite{jpdyn}. Thus hybrid anyons can form bound states
due to the possibility of magnetic trapping \cite{cs1,cs2}. At the quantum
level
the effect of these classical trapped states can be observable in the form
of resonances characterised by Landau levels. This also means that
when one constructs effective field theory for vortices like in
\cite{ezawa1,ezawa2}
the counterterm due to the short range statistical interaction can be
much smaller than expected. Due to the finite width of topological vortices
the statistical interaction is effectively switched off at short distances.
Also the internal structure of the hybrid anyon can manifest itself
by excitations with discrete energy levels.

$\;Aknowledgement.\;$ I would like to thank Igor Barashenkov for
discussions. This research was supported partially by the KBN grant
2P302 049 05 and by TEMPUS studentship which enabled my stay in Utrecht.

\end{document}